\documentclass[%
 aip,
 mph,%
 amsmath,amssymb,
 superscriptaddress,
 floatfix,
preprint,
eqsecnum,%
numerical,%
]{revtex4-2}

\usepackage{graphicx}
\DeclareGraphicsExtensions{.eps,.pdf,.png,.jpg}
\usepackage[caption=false
]{subfig}
\usepackage{epsfig}
\usepackage{epstopdf}
\usepackage{lipsum}
\usepackage{wrapfig}
\usepackage{makecell}
\usepackage{tabularx}
\usepackage{multirow}
\usepackage{booktabs}
\usepackage{lmodern}
\usepackage{longtable}
\usepackage{amsmath}
\usepackage[OT2,T1]{fontenc}
\usepackage[bookmarks=false]{hyperref}
\usepackage{color}

\hypersetup{pdfauthor={},pdfborder=0 0 0,colorlinks=true,citecolor=blue,linkcolor=blue,urlcolor=blue,}

\DeclareSymbolFont{cyrletters}{OT2}{wncyr}{m}{n}
\DeclareMathSymbol{\comb}{\mathalpha}{cyrletters}{"58}

\newcommand{\ben}{\begin{eqnarray}\displaystyle}
\newcommand{\een}{\end{eqnarray}}

\newcommand{\edit}{\textcolor{red}}

\usepackage{placeins}

\begin{document}
\title{Dual-energy cardiac DSA imaging using a dual-layer flat-panel detector: \textit{in vivo} validation study}
\author{Yuhang Tan}
\author{Jiongtao Zhu}
\author{Xin Zhang}
\affiliation{Research Center for Advanced Detection Materials and Medical Imaging Devices, Shenzhen Institutes of Advanced Technology, Chinese Academy of Sciences, Shenzhen, Guangdong 518055, China}

\author{Haodi Wu}
\affiliation{Research Center for Advanced Detection Materials and Medical Imaging Devices, Shenzhen Institutes of Advanced Technology, Chinese Academy of Sciences, Shenzhen, Guangdong 518055, China}
\affiliation{State Key Laboratory of Biomedical Imaging Science and System, Shenzhen, Guangdong 518055, China}

\author{Xiao Xuan}
\author{Mang Feng}
\author{Mingzhi Cui}
\affiliation{Beijing Wandong Medical Technology Co., Ltd., Beijing 100015, China}

\author{Hairong Zheng}
\affiliation{Paul C Lauterbur Research Center for Biomedical Imaging, Shenzhen Institutes of Advanced Technology, Chinese Academy of Sciences, Shenzhen, Guangdong 518055, China}
\affiliation{State Key Laboratory of Biomedical Imaging Science and System, Shenzhen, Guangdong 518055, China}

\author{Dong Liang}
\affiliation{Research Center for Medical Artificial Intelligence, Shenzhen Institutes of Advanced Technology, Chinese Academy of Sciences, Shenzhen, Guangdong 518055, China}
\affiliation{Paul C Lauterbur Research Center for Biomedical Imaging, Shenzhen Institutes of Advanced Technology, Chinese Academy of Sciences, Shenzhen, Guangdong 518055, China}
\affiliation{State Key Laboratory of Biomedical Imaging Science and System, Shenzhen, Guangdong 518055, China}

\author{Yongshuai Ge}
 \thanks{Scientific correspondence should be addressed to Yongshuai Ge (ys.ge@siat.ac.cn).}
\affiliation{Research Center for Advanced Detection Materials and Medical Imaging Devices, Shenzhen Institutes of Advanced Technology, Chinese Academy of Sciences, Shenzhen, Guangdong 518055, China}
\affiliation{Paul C Lauterbur Research Center for Biomedical Imaging, Shenzhen Institutes of Advanced Technology, Chinese Academy of Sciences, Shenzhen, Guangdong 518055, China}
\affiliation{State Key Laboratory of Biomedical Imaging Science and System, Shenzhen, Guangdong 518055, China}

\date{\today}

\begin{abstract}

{\bf Background:} The temporal misregistration in conventional X-ray digital subtraction angiography (DSA) remains a key challenge for its practical application in cardiac catheterization, leading to incomplete separation of cardiac vessels from background anatomical structures.\\
	\noindent
{\bf Purpose:} In this study, the X-ray spectral information is unveiled for DSA imaging to generate better visualization of cardiac vessels with negligible anatomical background.\\
	\noindent
{\bf Methods:} A dedicated dual-energy DSA imaging model is proposed based on the subtraction residuals between pre-contrast and post-contrast spectral projections, enabling accurate material decomposition without requiring precise scatter correction. Specifically, a dual-layer flat-panel detector (DL-FPD) is assumed to acquire the needed temporally registered low-energy and high-energy projections. Moreover, a penalized likelihood material decomposition algorithm is proposed to enhance the material decomposition performance. To validate, \textit{in vivo} swine experiments were conducted for the first time on a clinical DSA system equipped with a DL-FPD.\\
	\noindent
{\bf Results:} Monte Carlo simulation results demonstrated that the impact of scatter on the subtraction residuals can be reduced by more than 80\%, substantially improving material decomposition accuracy. Compared with analytical material decomposition, the penalized-likelihood material decomposition algorithm significantly improves image contrast-to-noise ratio (CNR), enabling the visualization of small vessels and even the myocardial blush effect. Furthermore, compared with conventional dual-energy subtraction, the penalized-likelihood approach provides more quantitative material maps with fewer overlapping anatomical background artifacts.\\
	\noindent
{\bf Conclusion:} This preliminary study verifies that dual-energy DSA imaging method is able to generate high-quality cardiovascular imaging results.\\
\end{abstract}


\maketitle

\section{Introduction}
\label{sec:introduction}
By far, digital subtraction angiography (DSA) remains the clinical gold standard for vascular mapping and minimally invasive interventional guidance\cite{chang1984dsa}. Its vital applications cross multiple medical areas such as neurovascular diseases\cite{davis2013dsa4d}, peripheral arterial diseases\cite{norgren2007tasc} and so on. The discovery of DSA technique\cite{mistretta1981digital} by Prof. Charles Mistretta and his collaborators at the University of Wisconsin-Madison in 1980s solved the distracting anatomical background overlapping problem by subtracting the pre-contrast X-ray image from subsequent the post-contrast images. As a consequence, only the iodine contrast medium filled vessels are isolated, making it much easier for physicians to accurately diagnose the vascular lesions.

For neurovascular applications, DSA deals with a rigid skull and a stationary brain with negligible intrinsic organ motion, therefore, the background bone can be nicely removed after subtraction, generating clear images of tiny capillaries and aneurysms. When come to the cardiovascular applications, in which the heart is constantly beating (cardiac motion) and the chest is constantly moving (respiratory motion), the traditional DSA fails to generate clear vessel images because of strong temporal misregistration between the pre-contrast and the post-contrast images\cite{meijering1999retrospective}. As a result, only the post-contrast X-ray images, known as fluoroscopy or cineangiography, are utilized for routine coronary arteries instead of the standard DSA. Notably, the post-contrast X-ray images suffer complex background structures such as the spine, ribs, and sternum remain fully visible as bright structures. To overcome, currently, a higher concentration of iodine contrast medium is injected, and proper angular views to keep the coronary arteries away from the spine and ribs are adjusted. Nevertheless, tiny blockages may be easily missed or hidden behind the complex anatomical structures. Additionally, cineangiography is not able to provide the myocardial blush maps, which indicate how well blood is actually penetrating the microscopic capillaries of the heart muscle\cite{vantHof1998angiographic}.

To address the issue of anatomical overlapping, energy subtraction, which processes X-ray projections acquired at multiple energy levels using specific weighting coefficients to suppress either bone or soft tissue, has been proposed\cite{brody1981hybrid}. Because pre-contrast images are no longer required, temporal misregistration artifacts can be completely eliminated. However, achieving a clean isolation of the iodine contrast agent without residual background anatomy remains challenging\cite{Niu2014IterativeID}. To improve, recent studies have been focused on quantitative material decomposition on dual-energy or spectral X-ray projections to isolate distinct basis maps for soft tissue, bone, and iodine. Various approaches can be utilized to acquire dual-energy or spectral X-ray projections, including fast tube-voltage switching \cite{mccollough2015dual}, multi-layer flat-panel detectors (FPDs) \cite{kim2022towards}, and energy-resolving photon counting detectors \cite{Schlomka2008ExperimentalFO}. Among these solutions, the multi-layer FPD approach is considered the most promising due to its ability to capture the required spectral projections simultaneously. In a typical dual-layer FPD, for example, the low-energy (LE) projection is captured by the top layer, while the high-energy (HE) projection is acquired by the bottom layer. Usually, the top CsI:Tl scintillator layer is thinner than the bottom layer, and a copper filter is inserted between them to enhance beam spectral separation. Several investigations of the heart vessel imaging have been reported based on either DL-FPD\cite{staahl2021performance} or triple-layer FPD (TL-FPD)\cite{jiang2023approaches}. For example, Shi \textit{et al.} experimentally demonstrated the feasibility of material-specific decomposition using a DL-FPD to improve the separation and visualization of cardiac vessels~\cite{shi2020characterization}. However, the decomposed coronary arteries in the porcine heart specimen remained overlapped with the background spine and rib structures. Gang \textit{et al.} further demonstrated that a model-based material estimation algorithm enabled high-quality three-material decomposition with a DL-FPD in the presence of soft-tissue motion using a numerical XCAT thorax phantom\cite{gang2023threematerial}. More recently, Jiang \textit{et al.} reported the feasibility of three-material (water/calcium/iodine) decomposition using a TL-FPD combined with a ResUNet-based decomposition method\cite{jiang2023approaches}. Note that the performance of the above three material decomposition approaches rely on accurate scatter correction\cite{shi2020characterization}, which is usually quite challenging in practice.

In this study, we propose a novel scatter-suppressed, material-specific decomposition framework based on temporally subtracted dual-energy projections, improving the feasibility of dual-energy DSA in practical imaging scenarios. First, the dual-energy pre-contrast projections are subtracted from the corresponding post-contrast projections. The resulting dual-energy projection residuals contain substantial misregistered background anatomical structures, predominantly soft tissue and cardiac structures, along with residual bony structures. Second, a dedicated two-material decomposition model for soft tissue and iodine is established for these dual-energy projection residuals. In this model, the bone residuals are assumed to be effectively removed through image registration between the pre-contrast and post-contrast projections. Third, a penalized likelihood material decomposition algorithm is proposed to perform the material decomposition. Finally, \textit{in vivo} porcine experiments were conducted in our laboratory on a modified clinical DSA system equipped with a DL-FPD to validate the proposed dual-energy cardiac DSA imaging framework.

The rest of this paper is organized below: Section \ref{sec: method} presents the dual-energy cardiac DSA imaging model, the penalized likelihood material decomposition method, the experimental setup and evaluation metrics. Section \ref{sec:results} presents the \textit{in vivo} porcine experimental results at different projection views. Section \ref{sec:conclusion} provides the discussions and a brief conclusion. 

\section{Method} \label{sec: method}
\subsection{Forward projection model}
If neglecting scatter, the X-ray signals recorded by the top and bottom layers of a DL-FPD can be expressed as
\begin{equation}
I_{\mathrm{LE}}=I_{0}\int_{\mathcal E} \Omega_{0}(E)Ee^{-\int_{\mathcal L }\mu(\vec{x},t,E)dl}(1-e^{-\mu_\mathrm{CsI}(E)L^{t}_\mathrm{CsI}})dE,
\label{eq:det1}
\end{equation}
and
\begin{eqnarray}
I_{\mathrm{HE}}=I_{0}\int_{\mathcal E} \Omega_{0}(E)Ee^{-\int_{\mathcal L }\mu(\vec{x},t,E)dl}e^{-\mu_\mathrm{CsI}(E)L^{t}_\mathrm{CsI}}e^{-\mu_\mathrm{flt}(E)L_\mathrm{flt}}(1-e^{-\mu_\mathrm{CsI}(E)L^{b}_\mathrm{CsI}})dE, \label{eq:det2}
\end{eqnarray}
correspondingly. Herein, $I_{0}$ denotes the incident X-ray intensity, $\Omega_{0}(E)$ denotes the incident X-ray beam spectrum, $\mu(\vec{x},t,E)$ denotes the object's time dependent linear attenuation coefficient, $\mu_\mathrm{CsI}(E)$ denotes the linear attenuation coefficient of CsI:TI scintillator, $\mu_\mathrm{flt}(E)$ denotes the linear attenuation coefficient of beam filter added between the top and bottom detector layers. Moreover, $L^{t}_\mathrm{CsI}$, $L_\mathrm{flt}$, $L^{b}_\mathrm{CsI}$ represent the thickness of the top detector layer, the beam filter, and the bottom detector layer, respectively.

By normalizing the recorded X-ray signals $I_{\mathrm{LE}}$ and $I_{\mathrm{HE}}$ by their corresponding sample-free signals $I_{0,\mathrm{LE}}$ and $I_{0,\mathrm{HE}}$, measured under the same exposure settings, the logarithmic projections are obtained as follows:
\begin{eqnarray}
\mathcal{P}_{LE}
=-\ln\!\left(\frac{I_{\mathrm{LE}}}
{I_{0,LE}}\right)=:\sum_{k}\mu_{k}(t,E_{k,LE})L_{k}, \\ \label{eq:W1}
\mathcal{P}_{HE}
=-\ln\!\left(\frac{I_{\mathrm{HE}}}
{I_{0,HE}}\right)=:\sum_{k}\mu_{k}(t,E_{k,HE})L_{k}. \label{eq:W2}
\end{eqnarray}
where $\mu_{k}(t,E)$ denotes the effective linear attenuation coefficient of the $k$-th basis material at energy $E$, $LE$ denotes low-energy, $HE$ denotes high-energy, $L_{k}$ corresponds to the length of the $k$-th basis material at time \edit{$t$} along the beam penetration path. For cardiac DSA imaging tasks, the default basis materials are soft-tissue, bone, and iodine.

\subsection{Dual-energy subtraction residuals}
In practice, $\mathcal{P}_{LE}$ and $\mathcal{P}_{HE}$ projections often suffer from severe scatter contamination, substantially degrading the material decomposition accuracy. To overcome, this work proposes to subtract the dual-energy pre-contrast projections from the corresponding post-contrast projections. Mathematically,
\begin{eqnarray}
\Delta\mathcal{P}_{LE} = \mathcal{P}^{post}_{LE}-\mathcal{P}^{pre}_{LE}, \label{eq:sd9} \\
\Delta\mathcal{P}_{HE} = \mathcal{P}^{post}_{HE}-\mathcal{P}^{pre}_{HE}, \label{eq:sd10}
\end{eqnarray}

The dual-energy pre-contrast projections are modeled as
\begin{eqnarray}
\mathcal{P}^{pre}_{LE}
&=:&\sum_{k\in\{B,T\}}\mu^{pre}_{k}(t^{pre},E_{k,LE}^{pre})L^{pre}_{k}, \label{eq:sd1} \\
\mathcal{P}^{pre}_{HE}
&=:&\sum_{k\in\{B,T\}}\mu^{pre}_{k}(t^{pre},E_{k,HE}^{pre})L^{pre}_{k}. \label{eq:sd2}
\end{eqnarray}
Here, the subscripts $B$, $T$, and $I$ denote bone, soft tissue, and iodine, respectively. No iodine is assumed to be present in the pre-contrast measurements.

The dual-energy post-contrast projections are similarly modeled as
\begin{eqnarray}
\mathcal{P}^{post}_{LE}
&=:&\sum_{k\in\{B,T,I\}}\mu^{post}_{k}(t^{post},E_{k,LE}^{post})L^{post}_{k}, \label{eq:sd3} \\
\mathcal{P}^{post}_{HE}
&=:&\sum_{k\in\{B,T,I\}}\mu^{post}_{k}(t^{post},E_{k,HE}^{post})L^{post}_{k}. \label{eq:sd4}
\end{eqnarray}

On these dual-energy subtraction residuals, the impact of scatter is substantially reduced, see Appendix for more details. Due to the continuous cardiac and respiratory motions, as a result, the dual-energy subtraction residuals would contain a certain amount of misregistered background anatomical structures, for example, soft tissue, iodine vessels, and bony structures. In the following discussions, it is assumed that the bony structures can be removed with proper image registration, namely,
\begin{eqnarray}
\mu^{pre}_{B}(t^{pre},E_{B,LE}^{pre})L^{pre}_{B}
=\mu^{post}_{B}(t^{post},E_{B,LE}^{post})L^{post}_{B}, \\ \label{eq:sd5}
\mu^{pre}_{B}(t^{pre},E_{B,HE}^{pre})L^{pre}_{B}
=\mu^{post}_{B}(t^{post},E_{B,HE}^{post})L^{post}_{B}, \label{eq:sd6}
\end{eqnarray}
As a result, at most two material components, soft tissue and iodine, need to be decomposed. Explicitly having,
\begin{eqnarray}
\Delta\mathcal{P}_{LE} = \mu^{post}_{T}(t^{post},E_{T,LE}^{post})L^{post}_{T}-\mu^{pre}_{T}(t^{pre},E_{T,LE}^{pre})L^{pre}_{T}+\mu^{post}_{I}(t^{post},E_{I,LE}^{post})L^{post}_{I}, \\ \label{eq:sd7}
\Delta\mathcal{P}_{HE} = \mu^{post}_{T}(t^{post},E_{T,HE}^{post})L^{post}_{T}-\mu^{pre}_{T}(t^{pre},E_{T,HE}^{pre})L^{pre}_{T}+\mu^{post}_{I}(t^{post},E_{I,HE}^{post})L^{post}_{I}. \label{eq:sd8}
\end{eqnarray}
Assuming that the energy dependent linear attenuation coefficient of soft-tissue before and after iodine contrast injection is similar, namely,
\begin{eqnarray}
\mu_{T}(E_{T, LE})=:\mu^{pre}_{T}(t^{pre},E_{T,LE}^{pre}) \approx \mu^{post}_{T}(t^{post},E_{T,LE}^{post}), \\
\mu_{T}(E_{T, HE})=:\mu^{pre}_{T}(t^{pre},E_{T,HE}^{pre}) \approx \mu^{post}_{T}(t^{post},E_{T,HE}^{post}).
\end{eqnarray}
Denoting $L_I=L_I^{post}$ and $\mu_I(E_{I,c})=\mu_I^{post}(t^{post},E_{I,c}^{post})$, with $c\in\{LE,HE\}$. The dual-energy subtraction residuals then become
\begin{eqnarray}
\label{eq:sd11}
\Delta\mathcal{P}_{LE} = \mu_{T}(E_{T, LE})(L^{post}_{T}-L^{pre}_{T})+\mu_{I}(E_{I, LE})L_{I}, \\ 
\label{eq:sd12}
\Delta\mathcal{P}_{HE} = \mu_{T}(E_{T, HE})(L^{post}_{T}-L^{pre}_{T})+\mu_{I}(E_{I, HE})L_{I}. 
\end{eqnarray}
Denoting $\Delta L_{T}=L^{post}_{T}-L^{pre}_{T}$, and taking the X-ray photon fluctuation into consideration, Eqs.~\eqref{eq:sd11}-\eqref{eq:sd12} can be re-formulated as
\begin{eqnarray}
\label{eq:sd13}
\Delta\mathcal{P}_{LE} = \mu_{T}(E_{T, LE})\Delta L_{T}+\mu_{I}(E_{I, LE})L_{I}+n_{LE}, \\ 
\label{eq:sd14}
\Delta\mathcal{P}_{HE} = \mu_{T}(E_{T, HE})\Delta L_{T}+\mu_{I}(E_{I, HE})L_{I}+n_{HE}. 
\end{eqnarray}
The Eqs.~\eqref{eq:sd13}-\eqref{eq:sd14} represent the dual-energy cardiac DSA imaging model utilized in this study. In practice, usually more than one single pre-contrast projection may be averaged to lower the image noise, namely,
\begin{eqnarray}
\label{eq:sd15}
\bar{\mathcal{P}}^{pre}_{LE}
=\frac{1}{N}\sum_{i=1}^{i=N}\mathcal{P}^{pre}_{LE}(t_{i}^{pre}), \\ 
\label{eq:sd16}
\bar{\mathcal{P}}^{pre}_{HE}
=\frac{1}{N}\sum_{i=1}^{i=N}\mathcal{P}^{pre}_{HE}(t_{i}^{pre}).
\end{eqnarray}
Herein, $N$ denotes the total number of pre-contrast projections.

\subsection{Penalized decomposition}
Analytical solutions of Eqs.~\eqref{eq:sd13}-\eqref{eq:sd14} may suffer from severe noise amplification. To avoid, iterative solutions that can suppress the noise amplification effect during material decomposition is carried out. Under an independent, zero-mean, equal-variance Gaussian noise approximation, the penalized least-squares objective is written as follows:
\begin{equation}
\hat{\boldsymbol{L}}=\arg\min_{\boldsymbol{L}}\ \frac{1}{2}\left\|\mathbf U\boldsymbol{L}-\Delta \boldsymbol{\mathcal P}\right\|_{2}^{2}+R(\boldsymbol{L}),
\label{eq:obj}
\end{equation}
where $\left\| \cdot \right\|_2$ denotes the ${l}_2$ norm. Independent smoothed total variation (TV) penalties are applied to the two materials:
\[
\begin{aligned}
R(\boldsymbol L)&=\beta_T\mathrm{TV}_{\varepsilon}(\Delta L_T)+\beta_I\mathrm{TV}_{\varepsilon}(L_I),\\
\mathrm{TV}_{\varepsilon}(u)&=\sum_j\sqrt{(D_xu)_j^2+(D_yu)_j^2+\varepsilon^2},
\end{aligned}
\]
where $D_x$ and $D_y$ denote spatial finite differences, $j$ indexes pixels, $\beta_T$ and $\beta_I$ are regularization weights, and $\varepsilon>0$ is a small smoothing parameter. In addition,
$$\boldsymbol{L}=\begin{bmatrix}\Delta L_{T}\\[2pt]
L_{I}\end{bmatrix}
$$
denotes the residual integration length of soft-tissue and iodine to be determined,
$$
\Delta \boldsymbol{\mathcal P}=\begin{bmatrix}\Delta \mathcal P_{LE}\\[2pt]
\Delta \mathcal P_{HE}\end{bmatrix}
$$
denotes the measured dual-energy subtraction residuals, and
$$
\mathbf U=\begin{bmatrix}
\mu_{T}(E_{T, LE}) & \mu_{I}(E_{I, LE})\\[2pt]
\mu_{T}(E_{T, HE}) & \mu_{I}(E_{I, HE})
\end{bmatrix}
\label{eq:xbM}
$$
denotes the attenuation coefficient matrix of soft-tissue and iodine pre-determined at low-energy and high-energy. The material maps and projection residuals are treated as vectorized images. The matrices $\mathbf U$ and $\boldsymbol\rho$ act on the two components at each pixel, and the norms are evaluated over all pixels. 

To solve, the alternating direction method of multipliers (ADMM) was employed\cite{boyd2011distributed}. Specifically, auxiliary variables are introduced to reformulate the original problem as an augmented Lagrangian (AL) form, which is then solved by alternately updating the primal variables, auxiliary variables, and Lagrange multipliers. By introducing an auxiliary constraint variable $z=\boldsymbol{L}$, the augmented Lagrangian of the objective function is rewritten as
\begin{eqnarray}
\mathcal{L}_{\edit{\boldsymbol{\rho}}}\!\left(\boldsymbol{L},\,z,\,\alpha\right)
=\frac{1}{2}\left\|\mathbf U\boldsymbol{L}-\Delta \boldsymbol{\mathcal P}\right\|_{2}^{2}
+R(z)+\big\langle \alpha,\,\boldsymbol{L}-z\big\rangle
+\frac{1}{2}\left\|\boldsymbol{L}-z\right\|_{\boldsymbol{\rho}}^{2},
\label{eq:al}
\end{eqnarray}
where $\alpha$ is the Lagrange multiplier (dual variable). The penalty matrix $\boldsymbol{\rho}=\operatorname{diag}(\rho_T,\rho_I)$ contains the positive penalty parameters for soft tissue and iodine, respectively, with $\|\boldsymbol v\|_{\boldsymbol{\rho}}^2=\rho_T\|v_T\|_2^2+\rho_I\|v_I\|_2^2$. The scaled multiplier is $\lambda=\boldsymbol{\rho}^{-1}\alpha$.The ADMM iterations consist of two minimization steps and a multiplier update:
\begin{equation}
\left\{\begin{aligned}
\boldsymbol{Z}^{n}: & \boldsymbol{z}^{n}=\arg\min_{\boldsymbol{z}}\;
R(\boldsymbol{z})+\frac{1}{2}\left\|\boldsymbol{L}^{n-1}-\boldsymbol{z}+\boldsymbol{\lambda}^{n-1}\right\|_{\boldsymbol{\rho}}^{2},\\[3pt]
\boldsymbol{X}^{n}: & \boldsymbol{L}^{n}=\arg\min_{\boldsymbol{L}}\;
\frac{1}{2}\left\|\mathbf U\boldsymbol{L}-\Delta \boldsymbol{\mathcal P}\right\|_{2}^{2}
+\frac{1}{2}\left\|\boldsymbol{L}-\boldsymbol{z}^{n}+\boldsymbol{\lambda}^{n-1}\right\|_{\boldsymbol{\rho}}^{2},\\[3pt]
\boldsymbol{\Lambda}^{n}: & \boldsymbol{\lambda}^{n}=\boldsymbol{\lambda}^{n-1}+\eta^{n}\left(\boldsymbol{L}^{n}-\boldsymbol{z}^{n}\right).
\end{aligned}\right.
\label{eq:admm}
\end{equation}
Specifically, the $\boldsymbol{Z}^{n}$ step applies TV regularization, the $\boldsymbol{X}^{n}$ step updates the two-material vector $\boldsymbol L^n$ by balancing data fidelity with the auxiliary constraint, and the $\boldsymbol{\Lambda}^{n}$ update promotes agreement between $\boldsymbol L$ and $\boldsymbol z$. Here, $n\in\{1,2,\dots,N_{\mathrm{iter}}\}$, $N_{\mathrm{iter}}$ is the maximum number of iterations, and $\eta^n$ is the multiplier update step size.

\subsection{Experiments}
\subsubsection{Numerical experiments}
To quantitatively assess the scatter-artifact suppression effect of the proposed subtraction operations in Eqs.~\eqref{eq:sd9} and \eqref{eq:sd10}, Monte Carlo (MC) simulations of a dual-layer flat-panel detector (DL-FPD) based DSA imaging system were conducted on the MC-GPU platform\cite{badal2009accelerating}. The spectral and geometric settings in MC simulations were identical to those of the physical experiments, as listed in Table~\ref{table1}. The detector array consisted of $400\times400$ pixels with a pixel pitch of 0.5~mm, corresponding to a $200\times200$~mm$^2$ field of view at the detector plane, and $6\times10^{11}$ photons were simulated per projection. The respiratory 4D CT data were obtained from the publicly available CT-vs-PET-VentilationImaging dataset~\cite{eslick2022ct}, and the cardiovascular structures were obtained from DeepSA dataset~\cite{zeng2024pretrained}. Projections corresponding to varied respiratory phases with and without iodine contrast were simulated. In total, four different signals were recorded: the primary (scatter-free) signal $I_\mathrm{p}$, and the scatter signal $I_\mathrm{s}$ comprising Compton, Rayleigh, and multi-scatter components.

\subsubsection{Physical experiments}
\begin{figure}[t]
\centerline{\includegraphics[width=1\textwidth]{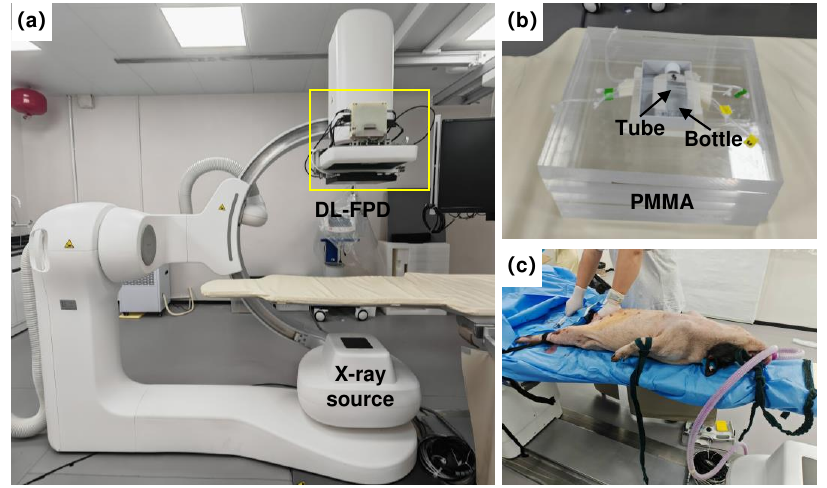}}
\caption{(a) The clinical DSA system equipped with the DL-FPD. (b) The quantitative phantom. (c) An anesthetized Bama pig used for the \textit{in vivo} angiography experiment.}\label{fig_experiment}
\end{figure}

Experiments were performed on a clinical DSA system (CGO-5200 PRO, Wandong Medical Technology Co., Ltd., Beijing, China) equipped with a dual-layer flat-panel detector (CareView 560RF-DE, CareRay Digital Medical Technology Co., Ltd., China), as shown in Fig.~\ref{fig_experiment}(a). Both detector layers have $1536\times1536$ pixels with a native pixel dimension of 0.154~mm $\times$ 0.154~mm, and the detector was operated in the $2\times2$ binning mode. An additional 0.3~mm copper interlayer was inserted between the top and bottom layers to enhance the spectral separation between the low-energy and high-energy projections. More detailed parameter settings are listed in Table~\ref{table1}. Due to the 6.6~mm physical gap between the top and bottom layers, a geometric registration was applied to align the low-energy and high-energy projections prior to subsequent processing. The sample-free low-energy and high-energy X-ray intensities were used to obtain the line-integrals.

{\renewcommand{\arraystretch}{1.0}
\begin{table}[htb]
\caption{Parameter settings used for experiments.}
\label{table1}
\centering
\begin{tabular}{cc}
\hline
Parameter & Value\\ \hline
Detector model            & CareView 560RF-DE \\ 
Tube voltage (kV)  & 120 \\ 
Pulsed tube current (mA)   & 400 \\ 
Pulse width (ms)   & 5 \\ 
Beam filtration     & 0.2~mm Cu \\ 
Interlayer filtration & 0.3~mm Cu \\ 
Dose per frame      & 460~$\mu$Gy/frame@60~cm \\ 
Frame rate (fps)    & 15 \\ 
SOD (mm) & 740 \\
SDD (mm) & 1045 \\
Detector array      & 1536$\times$1536 \\ 
Pixel dimension (mm$^2$) & 0.154$\times$0.154 \\ 
Detector binning    & 2$\times$2 \\ 
CsI:Tl thickness (mm) & 0.26 (top); 0.55 (bottom) \\
\hline
\end{tabular}
\end{table}

A self-made iodine solution phantom was used to quantitatively validate the proposed dual-energy material decomposition algorithm, as shown in Fig.~\ref{fig_experiment}(b). Iodine solutions with concentrations of 5, 2.5, and 1~mgI/ml were filled into a thin cubic bottle to mimic the myocardial blush. In addition, iodine solutions of 100 and 50~mgI/ml were filled into narrow plastic tubes with different diameters to mimic the coronary arteries. Moreover, PMMA slabs were stacked piecewise to mimic soft-tissue of 145 mm thick. The pure-water-filled bottle was first scanned to acquire the pre-contrast dual-energy projections, and the iodine-filled bottle and tubes were then acquired as the post-contrast dual-energy projections. The known iodine concentrations served as reference values. Note that slight displacements may appear due to the exchange of the water bottle and the iodine solution bottle.

An \textit{in vivo} animal angiography study was performed with a male Bama swine (8 months old, 27~kg), as shown in Fig.~\ref{fig_experiment}(c). The animal research protocols were reviewed and approved by the Institutional Animal Care and Use Committee (IACUC) of Shenzhen Advanced Medical Services Co., Ltd (Study No. AMS E2608012P). Iohexol was adopted as the contrast agent, whose stock solution had an iodine concentration of 300~mgI/ml and was diluted to approximately 50--100~mgI/ml before injection. The pre-contrast frames were first acquired without contrast injection, and the post-contrast frames were then acquired after injection. 

\section{Results} \label{sec:results}
\subsection{Numerical simulation results}
Numerical simulation results are shown in Fig.~\ref{fig_scatter}. Specifically, Figs.~\ref{fig_scatter}(a) and (b) show the low-energy pre- and post-contrast projections simulated at two distinct respiratory phases, respectively. Their difference, $\mathcal{P}_{LE}^{pre}-\mathcal{P}_{LE}^{post}$, is shown in Fig.~\ref{fig_scatter}(c), revealing the iodine-related signal together with residual anatomical structures caused by respiratory motion. Fig.~\ref{fig_scatter}(d) shows the probability density histograms of the pixel values in (a)--(c). The pre- and post-contrast projection distributions largely overlap, whereas the subtraction residuals are concentrated near zero, indicating substantial cancellation of the complicated anatomical background.

\begin{figure}[t]
\centerline{\includegraphics[width=1\textwidth]{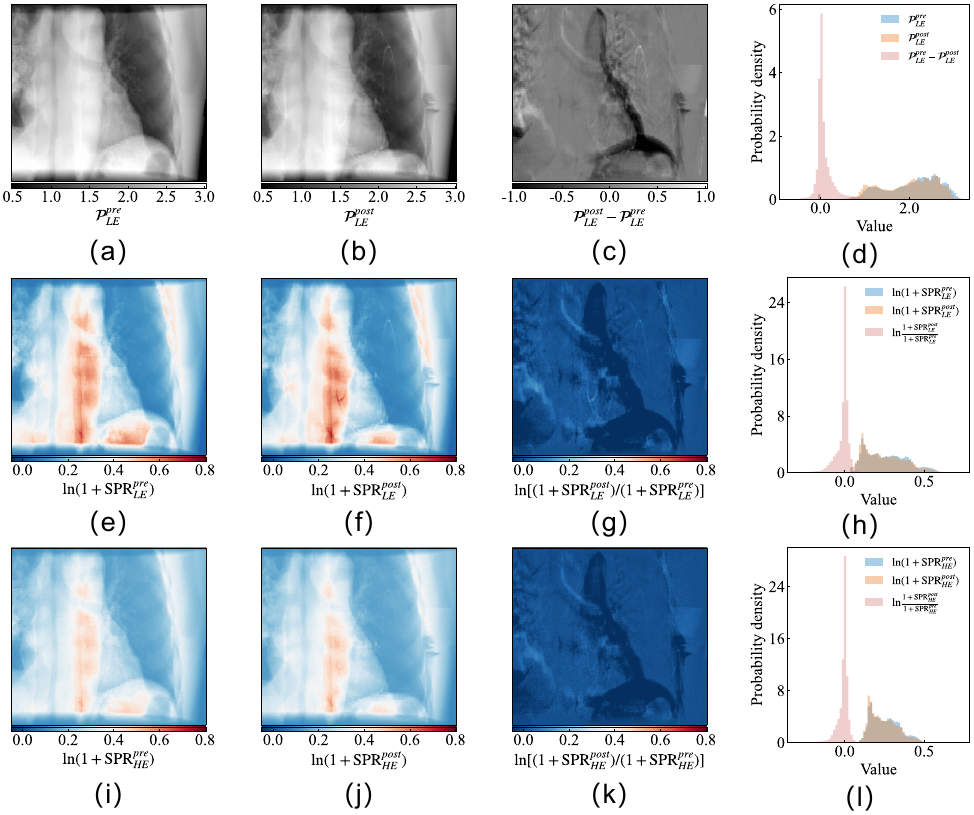}}
\caption{MC-simulated scatter results of the DL-FPD at two respiratory phases. (a),(b) Low-energy pre- and post-contrast projections, $\mathcal{P}_{LE}^{pre}$ and $\mathcal{P}_{LE}^{post}$, respectively. (c) The difference between (a) and (b), $\mathcal{P}_{LE}^{pre}-\mathcal{P}_{LE}^{post}$. (e),(f) Low-energy scatter contribution maps before subtraction, $\ln(1+\mathrm{SPR}_{LE}^{pre})$ and $\ln(1+\mathrm{SPR}_{LE}^{post})$, respectively. (g) The magnitude of the low-energy scatter residual after subtraction, $\ln[(1+\mathrm{SPR}_{LE}^{post})/(1+\mathrm{SPR}_{LE}^{pre})]$. (i)--(k) Corresponding high-energy scatter contribution maps before subtraction and residual magnitude after subtraction. (d),(h),(l) Probability density histograms of the pixel values in (a)--(c), (e)--(g), and (i)--(k), respectively.}\label{fig_scatter}
\end{figure}

As derived in the Appendix, the scatter contribution before subtraction is governed by $\ln(1+\mathrm{SPR})$, whereas the residual scatter contribution after subtraction depends on $\ln[(1+\mathrm{SPR}^{post})/(1+\mathrm{SPR}^{pre})]$. Accordingly, Figs.~\ref{fig_scatter}(e) and (f) show the low-energy scatter contribution maps, $\ln(1+\mathrm{SPR}_{LE}^{pre})$ and $\ln(1+\mathrm{SPR}_{LE}^{post})$, respectively, while Figs.~\ref{fig_scatter}(i) and (j) show the corresponding high-energy maps. The scatter contribution is higher in the low-energy projections than in the high-energy projections, while the pre- and post-contrast scatter distributions remain similar within each energy channel. Figs.~\ref{fig_scatter}(g) and (k) show the corresponding residual scatter maps after subtraction for the low- and high-energy channels, respectively, where the scatter contributions are substantially reduced. The probability density histograms in Figs.~\ref{fig_scatter}(h) and (l) further show that the pre- and post-contrast scatter distributions largely overlap, whereas the residual scatter distributions exhibit sharp peaks near zero. These results demonstrate that temporal subtraction suppresses the common scatter contribution in both energy channels. In these simulations, the scatter impact was reduced by more than 80\%, supporting the use of temporal subtraction to reduce scatter contamination prior to dual-energy material decomposition.

\subsection{Quantitative phantom results}
\begin{figure}[t]
\centerline{\includegraphics[width=0.85\textwidth]{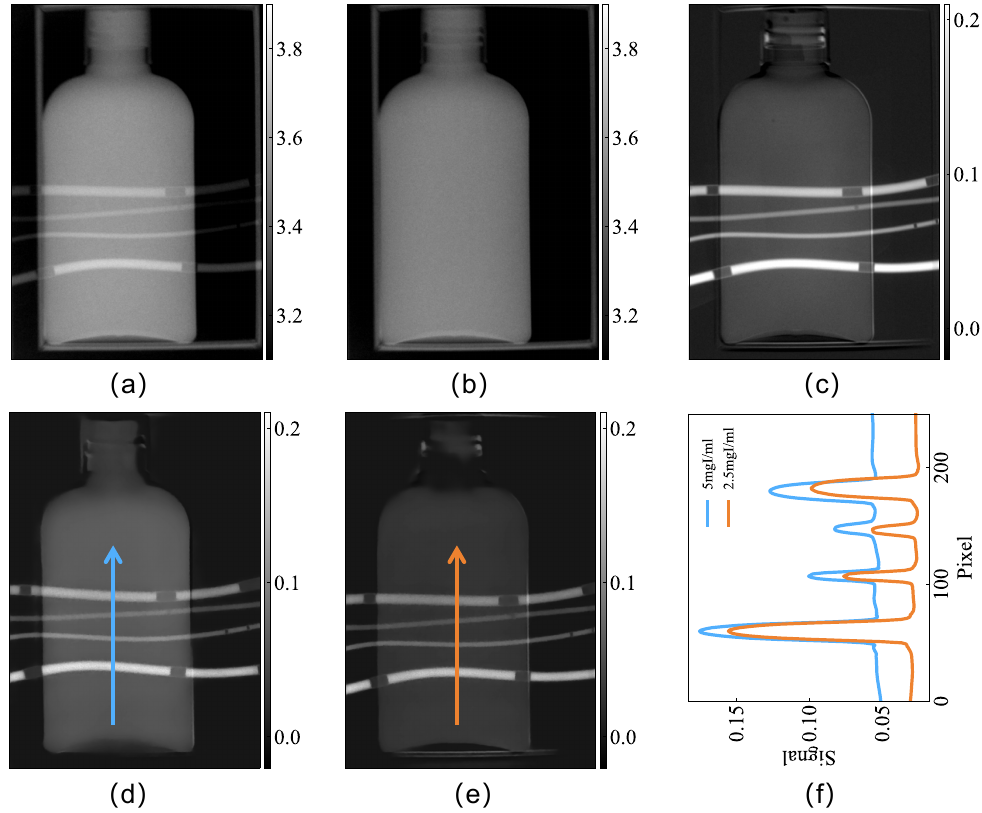}}
\caption{Quantitative validation results of the physical phantom. (a) Normalized projection of the 2.5~mgI/ml iodine-filled bottle and four vessel-mimicking tubes acquired with the PMMA background. The upper and lower pairs of tubes contain iodine solutions of 50 and 100~mgI/ml, respectively. (b) Normalized projection of the PMMA slabs and the water-filled bottle. (c) The residual projection for the 2.5~mgI/ml bottle. (d),(e) Iodine maps obtained by iterative material decomposition for bottle iodine concentrations of 5 and 2.5~mgI/ml, respectively. (f) Signal profiles along the blue and orange arrow directions in (d) and (e), respectively.}\label{fig_quantitative}
\end{figure}

Fig.~\ref{fig_quantitative} shows the quantitative results of the physical phantom.  Figs.~\ref{fig_quantitative}(a) and (b) show normalized signal and background projections acquired with 145~mm PMMA slabs, respectively. The signal contains the 2.5~mgI/ml iodine-filled bottle and four vessel-mimicking tubes, while the background contains the water-filled bottle. The average of 50 pre-contrast frames was subtracted from the signal frame in each energy channel to reduce noise. The resulting residual in Fig.~\ref{fig_quantitative}(c) retains the iodine-related signals while largely suppressing the common background. Figs.~\ref{fig_quantitative}(d) and (e) show the iodine maps obtained by the proposed penalized iterative decomposition method for bottle concentrations of 5 and 2.5~mgI/ml, respectively. The vessel-mimicking tubes are clearly resolved in both cases, and the bottle signal remains distinguishable at 2.5~mgI/ml over a thickness of 30~mm. The profiles in Fig.~\ref{fig_quantitative}(f) exhibit distinct vessel peaks above the bottle plateau, which is higher at 5~mgI/ml. The simultaneous visualization of vessel signals and a low-concentration iodine background suggests the potential of the proposed method for myocardial blush imaging.

\begin{figure}[t]
\centerline{\includegraphics[width=0.7\textwidth]{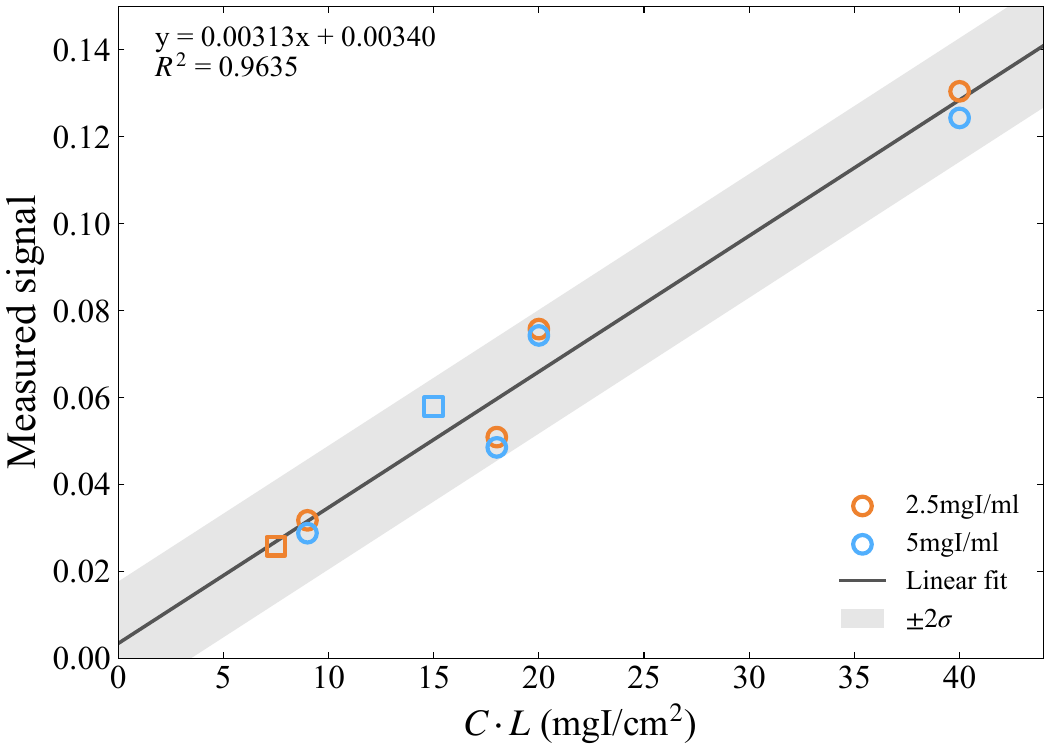}}
\caption{Relationship between the iodine concentration--thickness product ($C\cdot L$) and the decomposed iodine signal. Orange and blue squares represent bottle signals at iodine concentrations of 2.5 and 5~mgI/ml, respectively; circles of the same colors represent signals from the vessels overlying the corresponding bottles. The solid line shows the linear fit to all measurements ($R^{2}=0.9635$), and the shaded band indicates $\pm2$ standard deviations of the fitting residuals.}\label{fig_quantitative2}
\end{figure}

Fig.~\ref{fig_quantitative2} shows the relationship between the decomposed iodine signal and the iodine concentration--thickness product ($C\cdot L$). Bottle and vessel signals were quantified as plateau heights relative to baseline and baseline-subtracted Gaussian peak heights, respectively. A linear fit to all measurements yielded $y=0.00313x+0.00340$ with $R^{2}=0.9635$, where $x$ denotes $C\cdot L$ in mgI/cm$^{2}$ and $y$ denotes the measured iodine signal. Measurements from both bottle concentrations follow a similar trend, supporting an approximately linear response to iodine areal density across the tested bottle concentrations and vessel diameters.

\begin{figure}[t]
\centerline{\includegraphics[width=1.0\textwidth]{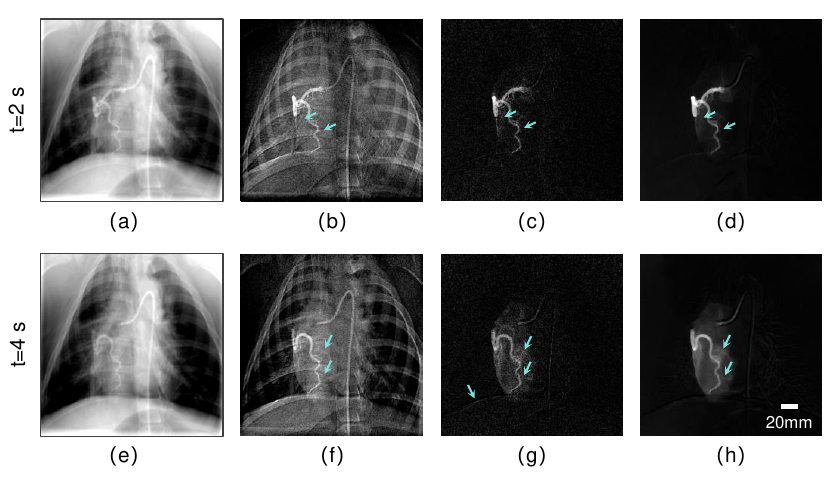}}
\caption{The \textit{in vivo} swine dual-energy imaging results acquired at a fixed projection view at $t=2$~s (top row) and $t=4$~s (bottom row). (a),(e) Normalized signal projections. (b),(f) Decomposition results obtained by the weighted subtraction method. (c),(g) Results obtained by the analytical decomposition method. (d),(h) Decomposition results obtained by the ADMM-based iterative method.}\label{fig_vivo1}
\end{figure}

\subsection{\textit{In vivo} swine results}
The \textit{in vivo} swine dual-energy imaging results acquired at a fixed projection view at $t=2$~s and $t=4$~s are shown in the top and bottom rows of Fig.~\ref{fig_vivo1}, respectively. The normalized projections at the two time points are shown in Figs.~\ref{fig_vivo1}(a) and (e), correspondingly. In this preliminary study, the post-contrast projections were registered to the pre-contrast projections to minimize the impact of cardiac and respiratory motions. The resulting dual-energy subtraction residuals were processed via three methods for comparison: weighted subtraction [Figs.~\ref{fig_vivo1}(b) and (f)], direct analytical dual-energy decomposition [Figs.~\ref{fig_vivo1}(c) and (g)], and the proposed ADMM-based penalized iterative material decomposition [Figs.~\ref{fig_vivo1}(d) and (h)]. Specifically, the weighted subtraction was performed as $\mathcal{P}_{LE}-\kappa\cdot\mathcal{P}_{HE}$ with an empirically tuned scaling factor $\kappa=1.10$.

As observed, for both time points, the analytical and weighted-subtraction results retain substantial noise, which limits the visibility of small vessel branches. The ADMM-based iterative method reduces this noise and improves the delineation of the vessels indicated by the arrows. Importantly, at $t=4$~s, the iterative result also shows a profound diffuse iodine signal surround the vessels, indicating potential for visualization of myocardial blush effect.

\section{Discussions and Conclusion} \label{sec:conclusion}
In this study, X-ray dual-energy imaging is investigated to mitigate the temporal misregistration in conventional X-ray digital subtraction angiography (DSA) in cardiac catheterization, which usually contains substantial anatomical background. To eliminate the dramatic impact of scatter in achieving accurate material decomposition, a dedicated dual-energy DSA imaging model is proposed based on the subtraction residuals between pre-contrast and post-contrast spectral projections. Moreover, the dual-layer flat-panel detector (DL-FPD) is utilized to acquire the needed temporally registered low-energy and high-energy projections. Finally, a penalized likelihood material decomposition algorithm is proposed to perform material decomposition. To validate, \textit{in vivo} swine experiments were conducted for the first time on a modified clinical DSA system equipped with a DL-FPD. Monte Carlo simulation results demonstrated that the impact of scatter on the subtraction residuals can be reduced by more than 80\%, substantially improving material decomposition accuracy. Compared with analytical material decomposition, the penalized-likelihood material decomposition algorithm significantly improves image contrast-to-noise ratio (CNR), enabling the visualization of small vessels and even the myocardial blush effect. Furthermore, compared with conventional dual-energy subtraction, the penalized-likelihood approach provides more quantitative material maps with fewer overlapping anatomical background artifacts.

It is found that the subtraction operation may not be able to completely eliminate the impact of scatter, particularly for large magnitude cardiac and respiratory motion displacements. As a result, the current study mainly focuses on validating the proposed method with pre-contrast and post-contrast spectral projections that have relatively small motion displacements. To compensate for large motion displacements, in the future, we will continue to optimize the penalized likelihood material decomposition algorithm to suppress the impact of potential residual scatter. Moreover, the current penalized likelihood material decomposition algorithm is proposed only for soft-tissue and iodine decomposition. In the future, multi-material, for instance, bone, soft-tissue and iodine, decomposition should be considered. Given the ill-posed nature of dual-energy multi-material decomposition, we propose to consider the convolutional neural network (CNN) based deep learning decomposition method. For instance, the deep-unrolling iterative framework\cite{su2022generalized}. With the powerful capability of CNN, it is believed that accurate multi-basis ($\ge3$) images would be generated even for large cardiac and respiratory motion displacements.

In conclusion, this work has presented a dual-energy DSA imaging framework that exploits temporally subtracted dual-energy projections for material-specific decomposition without requiring explicit scatter correction. Both numerical scatter simulations and physical phantom validations have demonstrated the effectiveness of the subtraction scheme, and the \textit{in vivo} swine experiments have further showcased the clinical feasibility of the proposed method.

\section*{Appendix}
\label{app:scatter}
When scatter is included, the recorded X-ray signal is $I_c=I_{\mathrm{p},c}+I_{\mathrm{s},c}$, where $I_{\mathrm{p},c}$ and $I_{\mathrm{s},c}$ denote the primary and scatter signals, respectively. Before subtraction, the projection can be expressed as
\begin{equation}
\mathcal{P}_c = \ln\frac{I_{0,c}}{I_{\mathrm{s},c}+I_{\mathrm{p},c}}
= \ln\frac{I_{0,c}}{I_{\mathrm{p},c}(1+\mathrm{SPR}_c)}
=\ln\frac{I_{0,c}}{I_{\mathrm{p},c}}-\ln(1+\mathrm{SPR}_c),
\label{app:eq:proj}
\end{equation}
Here, $c\in\{\mathrm{LE},\mathrm{HE}\}$ denotes the energy channel, $I_{0,c}$ is the corresponding sample-free signal, and $\mathrm{SPR}_c=I_{\mathrm{s},c}/I_{\mathrm{p},c}$. The term $\ln(I_{0,c}/I_{\mathrm{p},c})$ denotes the scatter-free projection.

After subtraction, the residual signal is expressed as
\begin{equation}
\Delta\mathcal{P}_c = \mathcal{P}^{post}_c-\mathcal{P}^{pre}_c
= \ln\frac{I^{pre}_c}{I^{post}_c}
= \ln\frac{I_{\mathrm{p},c}^{pre}}{I_{\mathrm{p},c}^{post}} -\ln\frac{1+\mathrm{SPR}^{post}_c}{1+\mathrm{SPR}^{pre}_c},
\label{app:eq:res1}
\end{equation}
where superscripts $pre$ and $post$ denote pre-contrast and post-contrast measurements, respectively. Unlike the scatter contribution term $\ln(1+\mathrm{SPR}_c)$ in Eq.~\eqref{app:eq:proj}, the scatter contribution now becomes $\ln[(1+\mathrm{SPR}^{post}_c)/(1+\mathrm{SPR}^{pre}_c)]$ and can be reduced when the pre-contrast and post-contrast SPRs are similar. As a consequence, the impact of scatter can be substantially suppressed after signal subtraction, see Fig.~\ref{fig_scatter}.
\section*{Acknowledgements}
The authors acknowledge the use of Doubao and ChatGPT for proofreading and grammar checking during manuscript
preparation.

\section*{Funding}
This work was supported in part by the National Key Research and Development Program (2024YFF0507800, 2024YFF0507804) and the National Natural Science Foundation of China (62422123, 12505357).

\bibliography{mybib}

@article{shi2020characterization,
  title={Characterization and potential applications of a dual-layer flat-panel detector},
  author={Shi, Linxi and Lu, Minghui and Bennett, N Robert and Shapiro, Edward and Zhang, Jin and Colbeth, Richard and Star-Lack, Josh and Wang, Adam S},
  journal={Medical physics},
  volume={47},
  number={8},
  pages={3332--3343},
  year={2020},
  publisher={Wiley Online Library}
}

@article{staahl2021performance,
  title={Performance characterization of a prototype dual-layer cone-beam computed tomography system},
  author={St{\aa}hl, Fredrik and Sch{\"a}fer, Dirk and Omar, Artur and van de Haar, Peter and van Nijnatten, Fred and Withagen, Paul and Thran, Axel and Hummel, Erik and Menser, Bernd and Holmberg, {\AA}ke and others},
  journal={Medical Physics},
  volume={48},
  number={11},
  pages={6740--6754},
  year={2021},
  publisher={Wiley Online Library}
}

@article{Schlomka2008ExperimentalFO,
  title={Experimental feasibility of multi-energy photon-counting K-edge imaging in pre-clinical computed tomography},
  author={Jens-Peter Schlomka and Ewald Roessl and Ralf Dorscheid and S Vaughn Dill and Gerhard Martens and Thomas Istel and Christian B{\"a}umer and Christoph Herrmann and Roger Steadman and G{\"u}nter Zeitler and Amir Livne and Roland Proksa},
  journal={Physics in Medicine \& Biology},
  year={2008},
  volume={53},
  number={15},
  pages={4031 - 4047},
  publisher={IOP Publishing}
}

@article{Niu2014IterativeID,
  title={Iterative image-domain decomposition for dual-energy CT},
  author={Niu, Tianye and Dong, Xue and Petrongolo, Michael and Zhu, Lei},
  journal={Medical physics},
  volume={41},
  number={4},
  pages={041901},
  year={2014},
  publisher={Wiley Online Library}
}

@article{eslick2022ct,
  title={Ct ventilation as a functional imaging modality for lung cancer radiotherapy (ct-vs-pet-ventilation-imaging)},
  author={Eslick, Enid M and Kipritidis, John and Gradinscak, Denis and Stevens, Mark J and Bailey, Dale L and Harris, Benjamin and Booth, Jeremy T and Keall, Paul J},
  journal={The Cancer Imaging Archive},
  year={2022}
}

@article{zeng2024pretrained,
  title={Pretrained subtraction and segmentation model for coronary angiograms},
  author={Zeng, Yunjie and Liu, Han and Hu, Juan and Zhao, Zhengbo and She, Qiang},
  journal={Scientific Reports},
  volume={14},
  number={1},
  pages={19888},
  year={2024},
  publisher={Nature Publishing Group UK, London}
}

@article{chang1984dsa,
  title={Digital subtraction angiography in interventional radiology},
  author={Chang, Richard and Kaufman, Scott L and Kadir, Shabnam and Mitchell, Stanley E and White, Robert I},
  journal={AJR. American Journal of Roentgenology},
  volume={142},
  number={2},
  pages={363--366},
  year={1984},
  doi={10.2214/ajr.142.2.363}
}

@article{davis2013dsa4d,
  title={4D digital subtraction angiography: implementation and demonstration of feasibility},
  author={Davis, Benjamin and Royalty, Kevin and Kowarschik, Markus and Rohkohl, Christopher and Oberstar, Evan and Aagaard-Kienitz, Beverly and Niemann, David and Ozker, Omer and Strother, Charles and Mistretta, Charles},
  journal={AJNR. American Journal of Neuroradiology},
  volume={34},
  number={10},
  pages={1914--1921},
  year={2013},
  doi={10.3174/ajnr.A3529}
}

@article{norgren2007tasc,
  title={Inter-Society Consensus for the Management of Peripheral Arterial Disease ({TASC II})},
  author={Norgren, Lars and Hiatt, William R and Dormandy, Jeffrey A and Nehler, Mark R and Harris, Kenneth A and Fowkes, F G R},
  journal={Journal of Vascular Surgery},
  volume={45},
  number={1, Supplement S},
  pages={S5--S67},
  year={2007},
  doi={10.1016/j.jvs.2006.12.037}
}

@article{mistretta1981digital,
  title={Digital angiography: a perspective},
  author={Mistretta, Charles A and Crummy, Andrew B and Strother, Charles M},
  journal={Radiology},
  volume={139},
  number={2},
  pages={273--276},
  year={1981},
  doi={10.1148/radiology.139.2.7012918}
}

@article{meijering1999retrospective,
  title={Retrospective motion correction in digital subtraction angiography: a review},
  author={Meijering, Erik H W and Niessen, Wiro J and Viergever, Max A},
  journal={IEEE Transactions on Medical Imaging},
  volume={18},
  number={1},
  pages={2--21},
  year={1999}
}

@article{brody1981hybrid,
  title={Hybrid subtraction for improved arteriography},
  author={Brody, William R},
  journal={Radiology},
  volume={141},
  number={3},
  pages={828--831},
  year={1981}
}

@article{mccollough2015dual,
  title={Dual- and multi-energy {CT}: principles, technical approaches, and clinical applications},
  author={McCollough, Cynthia H and Leng, Shuai and Yu, Lifeng and Fletcher, Joel G},
  journal={Radiology},
  volume={276},
  number={3},
  pages={637--653},
  year={2015},
  doi={10.1148/radiol.2015142631}
}

@article{kim2022towards,
  title={Towards subtraction angiography using a multi-layered {X}-ray detector},
  author={Kim, Jeehwan and Yoo, Seungryong and Kim, Junwoo and Kim, Ho Kyung},
  journal={Physica Medica},
  volume={95},
  pages={32--40},
  year={2022}
}

@article{jiang2023approaches,
  title={Approaches for three material decomposition using a triple-layer flat-panel detector},
  author={Jiang, Xiao and Stayman, J Webster and Gang, Grace J},
  journal={Proceedings of SPIE--the International Society for Optical Engineering},
  volume={12463},
  pages={124630X},
  year={2023},
  doi={10.1117/12.2654468}
}

@article{gang2023threematerial,
  title={Three-material decomposition using a dual-layer flat panel detector in the presence of soft tissue motion},
  author={Gang, Grace J and Stayman, J Webster},
  journal={Proceedings of SPIE--the International Society for Optical Engineering},
  volume={12463},
  pages={124630Y},
  year={2023},
  doi={10.1117/12.2654443}
}

@article{vantHof1998angiographic,
  title={Angiographic assessment of myocardial reperfusion in patients treated with primary angioplasty for acute myocardial infarction: myocardial blush grade},
  author={van 't Hof, Arnoud W J and Liem, Anhoar and Suryapranata, Harry and Hoorntje, Jan C A and de Boer, Menko-Jan and Zijlstra, Felix},
  journal={Circulation},
  volume={97},
  number={23},
  pages={2302--2306},
  year={1998},
  doi={10.1161/01.CIR.97.23.2302}
}

@article{boyd2011distributed,
  title={Distributed optimization and statistical learning via the alternating direction method of multipliers},
  author={Boyd, Stephen and Parikh, Neal and Chu, Eric and Peleato, Borja and Eckstein, Jonathan},
  journal={Foundations and Trends in Machine Learning},
  volume={3},
  number={1},
  pages={1--122},
  year={2011}
}

@article{badal2009accelerating,
  title={Accelerating {Monte Carlo} simulations of photon transport in a voxelized geometry using a massively parallel graphics processing unit},
  author={Badal, Andreu and Badano, Aldo},
  journal={Medical Physics},
  volume={36},
  number={11},
  pages={4878--4880},
  year={2009},
  doi={10.1118/1.3231824}
}

@Article{su2022generalized,
  author    = {Su, Ting and Cui, Zhuoxu and Yang, Jiecheng and Zhang, Yunxin and Liu, Jian and Zhu, Jiongtao and Gao, Xiang and Fang, Shibo and Zheng, Hairong and Ge, Yongshuai and others},
  journal   = {Physics in Medicine \& Biology},
  title     = {Generalized deep iterative reconstruction for sparse-view CT imaging},
  year      = {2022},
  number    = {2},
  pages     = {025005},
  volume    = {67},
  publisher = {IOP Publishing},
}
\bibliographystyle{unsrt} 

\end{document}